\documentclass[aps,showpacs,prd,twocolumn]{revtex4}
\usepackage{graphicx}
\usepackage{amssymb}
\usepackage{amsmath}
\usepackage{color}
\usepackage{float}

\begin{document}

\title{Analytical self-dual\ solutions in a nonstandard Yang-Mills-Higgs
scenario}
\author{R. Casana$^{1}$, M. M. Ferreira Jr.$^{1}$, E. da Hora$^{1}$ and C.
dos Santos$^{2}$.}
\affiliation{$^{1}${Departamento de F\'{\i}sica, Universidade Federal do Maranh\~{a}o,
65085-580, S\~{a}o Lu\'{\i}s, Maranh\~{a}o, Brazil.}\\
$^{2}${Centro de F\'{\i}sica e Departamento de F\'{\i}sica e Astronomia,
Faculdade de Ci\^{e}ncias da Universidade do Porto, 4169-007, Porto,
Portugal.}}

\begin{abstract}
We have found analytical self-dual solutions within the generalized
Yang-Mills-Higgs model introduced in Ref. \cite{pau}. Such solutions are
magnetic monopoles satisfying Bogomol'nyi-Prasad-Sommerfield (BPS) equations
and usual finite energy boundary conditions. Moreover, the new solutions are
classified in two different types according {to} their capability of
recovering (or not) the usual 't Hooft--Polyakov monopole. Finally, we
compare the profiles of the solutions we found with the standard ones, from
which we comment about the main features exhibited by the new configurations.%
\newline
\end{abstract}

\maketitle

\section{Introduction}

\label{Intro}

Configurations supporting a nontrivial topology have been intensively
studied in connection with many areas of physics \cite{n5}. In particular,
in the context of High Energy Physics such configurations are described as
the static solutions inherent to some classical field models, which are
supposed to be endowed with a symmetry breaking potential for the
self-interacting scalar-matter sector. Consequently, these topological
solutions usually arise as the result of a symmetry breaking or a phase
transition.

The most common topological configuration is the kink \cite{n0}, which
stands for the static solution inherent to a (1+1)-dimensional model
containing only one self-interacting real Higgs field. Regarding higher
dimensional models, other examples of topological structures include the
vortex \cite{n1} and the magnetic monopole \cite{n3}. While the vortices are
defined in (1+2)-dimensional gauge models, such as the Maxwell-Higgs theory,
the magnetic monopole solutions appear in a (1+3)-dimensional
non-Abelian-Higgs gauge scenario. Specifically, the monopoles arise as
well-behaved finite energy solutions in a SO(3) Yang-Mills-Higgs model,
representing the interaction between a gauge and a real scalar triplets \cite%
{Shnir}. In a very special situation (i.e., in the absence of the Higgs
potential), the monopole solution turns out as a BPS structure \cite{n4}
supported by a set of first-order differential equations whose analytical
solution is the 't Hooft-Polyakov monopole \cite{n3}.

During the last years, a new class of topological solutions, called
topological k-defects, has been intensively investigated in the context of
the field theories presenting modified dynamics (k-field theories). The idea
of noncanonical dynamics is inspired in string theories where it arises in a
natural way. Such models have been used\ in several distinct physical
scenarios, with interesting results involving studies of the accelerated
inflationary phase of the universe \cite{n8}, strong gravitational waves 
\cite{sgw}, tachyon matter \cite{tm}, dark matter \cite{dm}, and others \cite%
{o}. In this context, some of us have studied the self-dual frameworks
engendered by some k-field theories \cite{o1}. Such BPS k-configurations, in
general, have asymptotic behavior (when $r\rightarrow 0$ and for $%
r\rightarrow \infty $) similar as their standard counterparts. However, the
generalized dynamics can induce variations in the defect amplitude, in the
characteristic length, and in the profile shape. Additional investigations
regarding the topological k-structures and their main features can also be
found in\ Ref. \cite{o2}. Concerning the searching of BPS solitons in new
models, one has also considered generalized theories mimicking the usual
defect solutions, in the so called twinlike models \cite{twin}, which
provide the very same solutions obtained by the usual model taken as the
starting point.

In a recent letter \cite{pau}, some of us have introduced the self-dual
framework inherent to a generalized Yang-Mills-Higgs model whose
noncanonical self-dual solutions also constitute magnetic monopoles. At this
first moment, our attention was focussed in attaining numerical solutions.\
Now, one interesting question\ naturally arises about the existence of
generalized Yang-Mills-Higgs models endowed with analytical BPS monopole
solutions. The purpose of the present paper is to go further into this issue
by introducing some effective non-Abelian gauge models whose self-dual
equations can be\ analytically solved. Such models here considered are
divided into two different classes, according to their capability of
recovering or not the 't Hooft-Polyakov monopole solution.

In order to present our results, this letter is organized as follows: in the
next Section, we briefly review the unusual Yang-Mills-Higgs model studied
in Ref. \cite{pau}, including its self-dual structure, from which one gets
the generalized BPS equations to be investigated. In Sec. III, we achieve
the main goal of this work by introducing the\ aforecited non-Abelian
models, the corresponding exact self-dual solutions being explicitly
presented. Then, we depict the related analytical profiles, from which we
verify that the new solutions are well-behaved. Furthermore, we compare
these solutions with the usual 't Hooft-Polyakov analytical ones, commenting
on the main features of the nonstandard configurations. Finally, in Section
IV, we present our ending comments and perspectives regarding future
investigations.

\section{\textbf{The theoretical model}}

\label{general}

We begin by reviewing the nonstandard Yang-Mills-Higgs model introduced in
Ref. \cite{pau}, whose dimensionless Lagrangian density is%
\begin{equation}
\mathcal{L}=-\frac{g\left( \phi ^{a}\phi ^{a}\right) }{4}F_{\mu \nu
}^{b}F^{\mu \nu ,b}+\frac{f\left( \phi ^{a}\phi ^{a}\right) }{2}D_{\mu }\phi
^{b}D^{\mu }\phi ^{b}\text{.}  \label{1}
\end{equation}%
Here, $F_{\mu \nu }^{a}=\partial _{\mu }A_{\nu }^{a}-\partial _{\nu }A_{\mu
}^{a}+e\epsilon ^{abc}A_{\mu }^{b}A_{\nu }^{c}$ is the Yang-Mills field
strength tensor, $D_{\mu }\phi ^{a}=\partial _{\mu }\phi ^{a}+e\epsilon
^{abc}A_{\mu }^{b}\phi ^{c}$ stands for the non-Abelian covariant
derivative, and $\epsilon ^{abc}$ is the antisymmetric Levi-Civita symbol
(with $\epsilon ^{123}=1$). Moreover, $g\left( \phi ^{a}\phi ^{a}\right) $
and $f\left( \phi ^{a}\phi ^{a}\right) $ are positive arbitrary functions
which change the dynamics of the non-Abelian fields in an exotic way.

In this letter, we focus our attention on the spherically symmetric
configurations arising from Lagrangian (\ref{1}). In this sense, we look for
static solutions described by the standard \textit{Ansatz}%
\begin{equation}
\phi ^{a}=x^{a}\frac{H\left( r\right) }{r}\text{\ \ \ and \ \ }A_{0}^{a}=0%
\text{,}  \label{a}
\end{equation}%
\begin{equation}
A_{i}^{a}=\epsilon _{iak}x_{k}\frac{W\left( r\right) -1}{er^{2}}\text{,}
\label{b}
\end{equation}%
where $r^{2}=x^{a}x^{a}$. The functions $H\left( r\right) $ and $W\left(
r\right) $ are supposed to behave\ according the finite energy boundary
conditions%
\begin{equation}
H\left( 0\right) =0\text{ \ \ and \ \ }W\left( 0\right) =1\text{,}
\label{cc1}
\end{equation}%
\begin{equation}
H\left( \infty \right) =\mp 1\text{ \ \ and \ \ }W\left( \infty \right) =0%
\text{,}  \label{cc2}
\end{equation}%
which also guarantee the breaking of the SO(3) symmetry\ inherent to the
Lagrangian (\ref{1}).

In general, given the arbitrariness of the generalizing functions $f\left(
\phi ^{a}\phi ^{a}\right) $ and $g\left( \phi ^{a}\phi ^{a}\right) $, the
corresponding Euler-Lagrange equations for $H\left( r\right) $ and $W\left(
r\right) $ can be extremely hard to solve, even in the presence of suitable
boundary conditions. Notwithstanding, spite of the complicated scenario, the
system also admits\textbf{\ }finite energy BPS structures, i.e., legitimate
field configurations obtained as the solutions of a set of first order
differential equations. In this sense,  whereas the standard approach \cite%
{n4} states that the BPS equations arise by requiring the minimization of
the energy of the overall model, one has to consider the spherically
symmetric expression for the energy density inherent to the non-Abelian
Lagrangian (\ref{1}):%
\begin{eqnarray}
\varepsilon  &=&\frac{g}{e^{2}r^{2}}\left( \left( \frac{dW}{dr}\right) ^{2}+%
\frac{\left( 1-W^{2}\right) ^{2}}{2r^{2}}\right)   \notag \\
&&+f\left( \frac{1}{2}\left( \frac{dH}{dr}\right) ^{2}+\left( \frac{HW}{r}%
\right) ^{2}\right) \text{.}  \label{ed}
\end{eqnarray}%
Here, it is worthwhile to point out that, as already verified in Ref. \cite%
{pau}, the self-duality of the resulting model only holds when one imposes
the following constraint:%
\begin{equation}
g=\frac{1}{f}\text{.}  \label{v}
\end{equation}%
In more details, taking Eq. (\ref{v}) into account, Eq. (\ref{ed}) can be
written in the form%
\begin{align}
\varepsilon & =\frac{f}{2}\left( \frac{dH}{dr}\pm \frac{1-W^{2}}{er^{2}f}%
\right) ^{2}+\frac{1}{e^{2}r^{2}f}\left( \frac{dW}{dr}\mp efHW\right) ^{2} 
\notag \\
& \mp \frac{1}{er^{2}}\frac{d}{dr}\left( H\left( 1-W^{2}\right) \right) 
\text{.}
\end{align}%
The minimization procedure then leads to the first-order equations%
\begin{equation}
\frac{dH}{dr}=\mp \frac{1-W^{2}}{er^{2}f}\text{,}  \label{bps1}
\end{equation}%
\begin{equation}
\frac{dW}{dr}=\pm efHW\text{,}  \label{bps2}
\end{equation}%
which are the BPS\ equations of the model. Therefore, the energy density of
the BPS states is%
\begin{equation}
\varepsilon _{bps}=\mp \frac{1}{er^{2}}\frac{d}{dr}\left( H\left(
1-W^{2}\right) \right) \text{,}  \label{buc}
\end{equation}%
while the total energy is reduced to%
\begin{equation}
E_{bps}=4\pi \int r^{2}\varepsilon _{bps}dr=\frac{4\pi }{e}\text{,}
\label{te}
\end{equation}%
whenever (\ref{cc1}) and (\ref{cc2}) are satisfied.

In Ref. (\ref{1}), for a specific choice of $f$, the BPS equations (\ref%
{bps1}) and (\ref{bps2}) were numerically solved fulfilling the finite
energy boundary conditions (\ref{cc1}) and (\ref{cc2}). The attained
profiles describe BPS magnetic monopole solutions with total energy given by
Eq. (\ref{te}) within the nonstandard Yang-Mills-Higgs scenario (\ref{1}).
In the next Section, we will deal with the attainment of analytical solution
for such a generalized model.


\section{Analytical solutions}

In this Section, we accomplish the main goal of this work by introducing
some effective models for which the BPS equations (\ref{bps1}) and (\ref%
{bps2}) can be solved analytically. providing well-behaved solutions endowed
with finite energy\textbf{.} Furthermore, we depict the corresponding
profiles choosing $e=1$\ and considering only the lower signs in Eqs. (\ref%
{cc2}), (\ref{bps1}), (\ref{bps2}) and (\ref{buc}). We also determine the
profiles for the BPS energy density (\ref{buc}) and for $r^{2}\varepsilon
_{bps}$ (the integrand of Eq. (\ref{te})). Then, by comparing the new
solutions and standard (analytical) one, we comment on the main features of
the generalized monopoles here presented.

Firstly, it is important to note that the usual Yang-Mills-Higgs scenario is
easily recovered by setting $f=1$, for which the BPS equations generate the
well-known 't Hooft-Polyakov analytical solution (already written in
accordance with our conventions):%
\begin{equation}
H_{_{tHP}}\left( r\right) =\frac{1}{\tanh r}-\frac{1}{r}\text{,}
\label{u1aa}
\end{equation}%
\begin{equation}
W_{_{tHP}}\left( r\right) =\frac{r}{\sinh r}\text{.}  \label{u1b}
\end{equation}

In the sequel, we present the profiles of the new solutions and we make a
comparison between them and the 't Hooft-Polyakov monopole solution,
commenting about the main features and differences among them (see Figs. 1,
2, 3 and 4 below). The noncanonical models to be examined in this letter are
divided into two different classes. The first class is related to those
models recovering the usual 't Hooft-Polyakov result (given an appropriated
limit), while the second one includes the models which do not. All these
solutions fulfill the finite energy boundary conditions, as expected.

Here, in order to introduce our results, we first point out that the BPS
equations (\ref{bps1}) and (\ref{bps2}) can be combined into a single
equation, i.e.,%
\begin{equation}
\frac{dW}{dr}\frac{dH}{dr}=\frac{\left( W^{2}-1\right) HW}{r^{2}}\text{.}
\label{ke}
\end{equation}%
which relates the solution for $H\left( r\right) $ to that for $W\left(
r\right) $. In this sense, for a given $H\left( r\right) $, Eq. (\ref{ke})
can be integrated to give the corresponding solution for $W\left( r\right) $%
, and vice-versa. Here, it is worthwhile to note that such strategy can be
used even to describe nonphysical scenarios, i.e., those for which $H\left(
r\right) $ and/or $W\left( r\right) $ dot not behave as (\ref{cc1}) and (\ref%
{cc2}).

In this work, as we are interested in the physical solutions only, we adopt
the following prescription: firstly, we choose an analytical solution for $%
H\left( r\right) $ satisfying the boundary conditions (\ref{cc1}) and (\ref%
{cc2}). Then, we calculate the corresponding solution for $W\left( r\right) $
by integrating Eq. (\ref{ke}) explicitly (as the reader can verify, the
solutions we have found this way automatically obey (\ref{cc1}) and (\ref%
{cc2})). A posteriori, we use such expressions to attain the one for $%
f\left( r\right) $ via the BPS equations (\ref{bps1}) and (\ref{bps2}).
Moreover, we also depict the corresponding exact profiles for the BPS energy
density Eq. (\ref{buc}) and for $r^{2}\varepsilon _{bps}$. Here, it is
important to say that all the solutions we have obtained for $f\left(
r\right) $ and $\varepsilon _{bps}$ are positive, as desired; see Eq. (\ref%
{buc}). In addition, all the noncanonical scenarios we have discovered
exhibit the very same total energy, i.e., $E_{bps}=4\pi $; see Eq. (\ref{te}%
).

At the first moment, the question about the generalization of the usual 't
Hooft--Polyakov solutions (\ref{u1aa}) and (\ref{u1b}) arises in a rather
natural way. Indeed, we have verified that such generalization is possible,
the resulting model belonging to the first class. In this sense, taking%
\begin{equation}
H\left( r\right) =\frac{1}{\tanh r}-\frac{1}{r}\text{,}  \label{sca}
\end{equation}
Eq. (\ref{ke}) can be integrated to attain%
\begin{equation}
W\left( r\right) =\frac{\sqrt{1-C_{1}}r}{\sqrt{\sinh ^{2}r-C_{1}r^{2}}}\text{%
,}  \label{scb}
\end{equation}%
where $C_{1}$ stands for a real constant such that $C_{1}<1$ (note that $%
C_{1}=0$ leads us back to the standard theory). In addition, using (\ref{sca}%
) and (\ref{scb}), eqs. (\ref{bps1}) and (\ref{bps2}) can be solved for $%
f\left( r\right) $, the result being%
\begin{equation}
f\left( r\right) =\frac{\sinh ^{2}r}{\sinh ^{2}r-C_{1}r^{2}}\text{.}
\label{wc}
\end{equation}%
Here, despite the noncanonical form of (\ref{wc}), the solution for $H\left( 
{r}\right) $,\ given in Eq. (\ref{sca}), is the same\textbf{\ }one of the
usual scenario (see Eq. (\ref{u1aa})). On the other hand, the solution for $%
W\left( {r}\right) $ exhibits a generalized structure, which reduces to Eq. (%
\ref{u1b}) when $C_{1}=0$.

Now, we use our prescription to introduce two examples of effective models
belonging to the second class, i.e., standing for new families of analytical
monopole solutions. Here, in order to define the first family, we choose the
analytical solution for $H\left( r\right) $ as%
\begin{equation}
H\left( r\right) =\frac{r}{r+1}\text{,}  \label{ssaa}
\end{equation}%
which indeed obeys the boundary conditions (\ref{cc1}) and (\ref{cc2}). In
the sequel, by solving Eq. (\ref{ke}), one gets that the corresponding
nonusual profile for $W\left( r\right) $ is%
\begin{equation}
W\left( r\right) =\frac{1}{\sqrt{1+C_{2}r^{2}e^{2r}}}\text{,}  \label{ssbb}
\end{equation}%
which also behaves according (\ref{cc1}) and (\ref{cc2}), $C_{2}$ being a
positive real constant. Furthermore, taking (\ref{ssaa}) and (\ref{ssbb})
into account, the self-dual equations (\ref{bps1}) and (\ref{bps2}) give%
\begin{equation}
f\left( r\right) =\frac{C_{2}\left( r+1\right) ^{2}e^{2r}}{1+C_{2}r^{2}e^{2r}%
}\text{.}  \label{ws}
\end{equation}%
In this case, we note that one has $f\neq 1$\ for any value of $C_{2}$. This
explains why the solutions (\ref{ssaa}) and (\ref{ssbb}) are always
different from the usual ones, (\ref{u1aa}) and (\ref{u1b}), respectively.

The last model to be studied is a little bit more sophisticated than the
previous ones. Even in this case, one still gets well-behaved solutions
which support the model itself.\ In this sense, the second family of models,
which do not recover the usual 't Hooft-Polyakov solution, is defined by%
\begin{equation}
H\left( r\right) =e^{h\left( r\right) }\text{.}  \label{sha}
\end{equation}%
Here, $h\left( r\right) $ is given by%
\begin{equation}
h\left( r\right) \equiv \frac{e^{-2r}\left( r^{2}+4r+1\right) -2r-1-2r^{2}%
\mbox{Ei}\left( 1,2r\right) }{2r^{2}},
\end{equation}%
with the function $\mbox{Ei}\left( 1,r\right) $ being the exponential
integral%
\begin{equation}
\mbox{Ei}\left( 1,r\right) \equiv \overset{\infty }{\underset{1}{\int }}%
\frac{e^{-rx}}{x}dx\text{.}
\end{equation}%
Also in this case, and despite the highly nonlinear structure of $H\left(
r\right) $, Eq. (\ref{ke}) still can be integrated exactly, the result being
a relatively simple analytical expression for $W\left( r\right) $, i.e.,%
\begin{equation}
W\left( r\right) =\frac{\sqrt{C_{3}+1}\left( r+1\right) }{\sqrt{C_{3}\left(
r+1\right) ^{2}+e^{2r}}}\text{,}  \label{shb}
\end{equation}%
where $C_{3}$ is a real constant such that $C_{3}>-1$. In addition, eqs. (%
\ref{bps1}) and (\ref{bps2}) give%
\begin{equation}
f\left( r\right) =\frac{re^{2r-h\left( r\right) }}{\left( r+1\right) \left(
C_{3}\left( r+1\right) ^{2}+e^{2r}\right) }\text{.}  \label{wh}
\end{equation}%
Moreover, one clearly see that both (\ref{sha}) and (\ref{shb}) behave
according (\ref{cc1}) and (\ref{cc2}), as expected.

Now, once we have introduced the noncanonical solutions for $H\left(
r\right) $ and $W\left( r\right) $, we compare their profiles by plotting
them in Figs. 1 and 2. Also, in Figs. 3 and 4, we show the corresponding
solutions for $\varepsilon _{bps}$ and $r^{2}\varepsilon _{bps}$,
respectively. The standard results are also shown, for comparison. In what
follows, we choose $C_{1}=C_{2}=C_{3}=0.5$. 
\begin{figure}[tbp]
\centering\includegraphics[width=8.5cm]{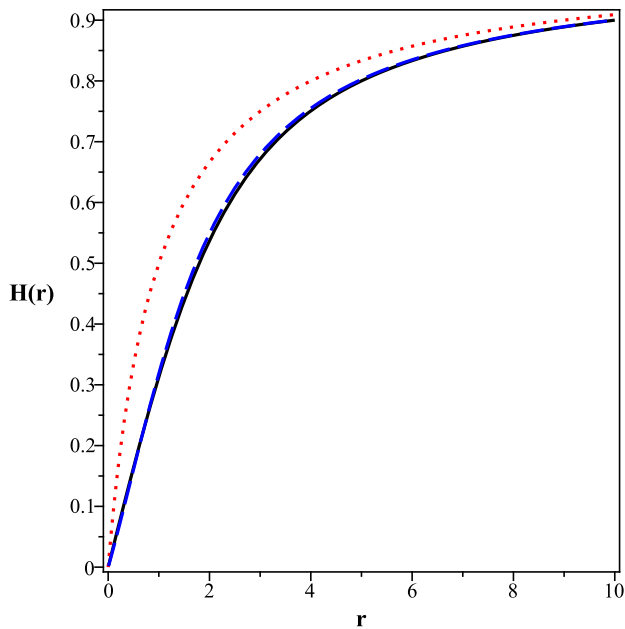}
\par
\vspace{-0.3cm}
\caption{Solutions to $H\left( r\right) $ given by {(\protect\ref{u1aa})
(usual case, solid black line), (\protect\ref{ssaa}) (dotted red line), and (%
\protect\ref{sha}) (dashed blue line). Here, (\protect\ref{sca}) mimics the
standard result}.}
\end{figure}

In Fig. 1, we depict the analytical solutions for $H\left( r\right) $. \ The
solution Eq. (\ref{ssaa}) is shown as a dotted red line, the dashed blue
line standing for solution Eq. (\ref{sha}). The usual profile, that of Eq. (%
\ref{u1aa}), is also shown (solid black line). It describes the solution
inherent to the choice (\ref{wc}), given by Eq. (\ref{sca}). One also notes
how close is the solution (\ref{sha}) to the standard profile. The overall
conclusion is that the solutions behave in the same general way: starting
from zero at the origin, they monotonically reach the asymptotic condition
in the limit $r\rightarrow \infty $. 
\begin{figure}[tbp]
\centering\includegraphics[width=8.5cm]{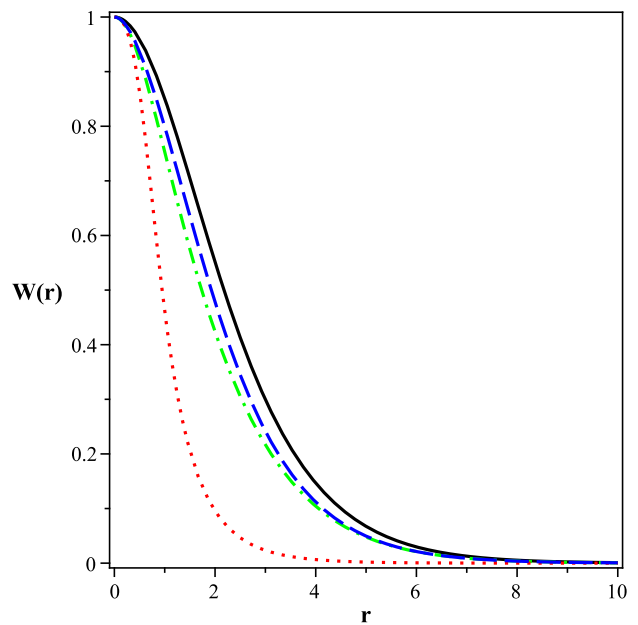}
\par
\vspace{-0.3cm}
\caption{Solutions to $W\left( r\right) $. Conventions as in FIG. 1. Here, {(%
\protect\ref{scb}) is represented by the dash-dotted green line.}}
\end{figure}

In Fig. 2, we show the solutions for $W\left( r\right) $, with Eq. (\ref{scb}%
) being represented by the dash-dotted green line. Also in this case, the
solutions have the same general profile, being lumps centered at the origin,
monotonically decreasing, and vanishing in the asymptotic limit. A
comparison\ between the new solutions with the standard one reveals that the
first ones present smaller characteristic lengths, the solution coming from
the choice (\ref{ws})\ being the one with the smallest range. 
\begin{figure}[tbp]
\centering\includegraphics[width=8.5cm]{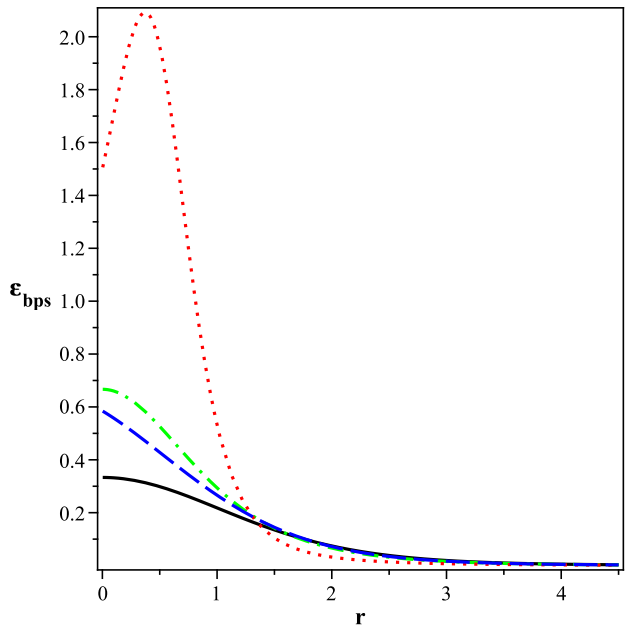}
\par
\vspace{-0.3cm}
\caption{Solutions to $\protect\varepsilon _{bps}$. Conventions as in the
previous figures.}
\end{figure}

The BPS energy density $\varepsilon _{bps}$ of the analytical solutions are
plotted in Fig. 3.\ The profiles related to the choices (\ref{wc}) and (\ref%
{wh}) behave as the standard one, that is, as a lump centered at $r=0$. On
the other hand, the solution inherent to (\ref{ws}) reaches its maximum
value at some finite distance $R$ from the origin, implying a ringlike
energy distribution in the plane. In addition, the solutions vanish
asymptotically,\ since the condition $\varepsilon _{bps}\left( r\rightarrow
\infty \right) \rightarrow 0$\ arises in a rather natural way from the
boundary conditions (\ref{cc2}). 
\begin{figure}[tbp]
\centering\includegraphics[width=8.5cm]{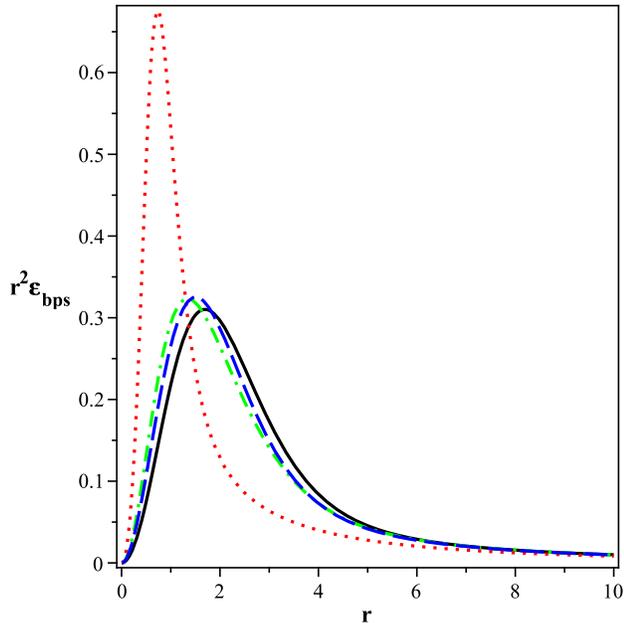}
\par
\vspace{-0.3cm}
\caption{Solutions to $r^{2}\protect\varepsilon _{bps}$. Conventions as in
the previous figures.}
\end{figure}

Finally, in Fig. 4\ we present the profiles for $r^{2}\varepsilon _{bps}$.
It clearly depicts a compensatory effect related to the profiles already
discussed in\ Ref. \cite{pau}: different solutions enclose the same area
(equal to the unity, according our conventions). As a consequence, the
resulting configurations have the very same total energy, given by $%
E_{bps}=4\pi $.


\section{Ending comments}

\label{end}

In this letter, we have extended a previous work \cite{pau} by\ introducing
non-Abelian effective models for which the resulting BPS equations can be
solved analytically.\ The starting point of such investigation was the
first-order formalism developed within a nonstandard Yang-Mills-Higgs theory 
\cite{pau}, whose dynamic is controlled by two positive generalizing
functions, $g\left( \phi ^{a}\phi ^{a}\right) $ and $f\left( \phi ^{a}\phi
^{a}\right) $. The non-Abelian fields were supposed to be described by the
standard spherically symmetric \textit{Ansatz} eqs. (\ref{a}) and (\ref{b}),
where the functions $H\left( r\right) $ and $W\left( r\right) $\ must behave
according the finite energy boundary conditions,\textbf{\ }(\ref{cc1}) and (%
\ref{cc2}). Our goal was to introduce effective Yang-Mills-Higgs models
whose corresponding BPS equations yield analytical solutions. Here, the
nonstandard models were divided into two different classes: the ones which
do recover the usual 't Hooft-Polyakov\textbf{\ }result\ (given the
appropriate limit), and the ones which do not; the last ones standing for
new families of analytical monopole solutions.

The profiles of the new solutions were depicted in Figs. 1, 2, 3 and 4. The
overall conclusion is that the effective models provide consistent and
well-behaved self-dual solutions which strongly support the models
themselves. Moreover, we have identified a particular family of nonstandard
models for which the BPS energy density exhibits a different profile (see
Fig. 4),\ with a ringlike energy distribution centered at $r\neq 0$. Thus,
we have shown that starting from a generalized Yang-Mills-Higgs framework we
can attain analytical self-dual solutions for non-Abelian magnetic monopoles.

Regarding future investigations, an interesting issue is the search for an
analytical description for the non-charged BPS vortices arising in the
generalized Maxwell-Higgs model proposed in Ref. \cite{bpsmh}. Moreover,
taking as the starting point the solutions we have presented in this work,
we intend to generate new self-dual profiles based on an appropriate
deformation prescription \cite{losano}. These issues are now under
consideration, with expected interesting results for a future contribution.

We thank CAPES, CNPq and FAPEMA (Brazilian agencies) for partial financial
support.

\end{document}